\documentclass{article}[10pt]

\usepackage{amsmath,amsthm, amssymb, latexsym}

\usepackage[english]{babel}

\usepackage{algorithmic}
\usepackage{algorithm}

\usepackage{verbatim}

\title{Factorization of Non-Commutative Polynomials}
\author{Fabrizio Caruso}

\newtheorem{remark}{Remark}
\newtheorem{example}{Example}

\begin{document}
\maketitle

\begin{abstract}
We describe an algorithm for the factorization of non-commutative polynomials 
over a field. The first sketch of this algorithm appeared in an
unpublished manuscript (literally hand written notes) by 
James~H.~Davenport more than 20 years ago.
This version of the algorithm 
contains some improvements with respect to the original sketch.
An improved version of the algorithm   
has been fully implemented 
in the \texttt{Axiom} computer algebra system.
\end{abstract}

\section{Introduction}
We describe an algorithm for the factorization of non-commutative
polynomials over a field, the first version of which
was described but never published by James~H. Davenport more than 20 years ago.
He wrote these notes \cite{Davenport} on the occasion of a 
visit to Pisa for a series of lectures and later forgot them.
These (hand-written) notes have been kept by Teo~Mora (Univ.~of~Genoa) who passed
them to Carlo~Traverso (Univ.~of~Pisa) who passed them to me.

The main ideas of the original algorithm and an application 
to cryptanalysis have been treated in \cite{CCT}, 
where we have shown
how to construct an attack on the non-commutative Polly-Crackers
 \cite{PollyCracker}
proposed in \cite{Rai}.

We are considering a free $K$-algebra
over a field $K$ for which effective polynomial system solving
is possible. We do not treat the cases
where some algebraic relations on the elements are imposed,
such as for example the case of linear differential operators
(\cite{Bronstein}, \cite{vanderPutSinger}).
The problem of factorizing a non-commutative polynomial
in this setting is clearly solvable by a brute-force
approach. Our algorithm if applied to univariate
polynomials coincides with the brute-force approach,
but it performs much better
in the multivariate case.

We give the full details of the algorithm,
two improvements of the original algorithm and 
we present its complete implementation in the
\texttt{Axiom} computer algebra system.
The implementation will be submitted to the \texttt{Axiom}
maintainers for inclusion in the next version of the system.

In order to simplify the description we will only 
consider the problem of factorizing a polynomial
in two factors of given total degrees.

\section{The Homogeneous Case}
The homogeneous case is a very special case.
The algorithm is simple and its correctness is
self-evident: 
we construct the two factors of a given polynomial $F$
by selecting a monomial $m=m_1 m_2$ and taking
the sum of the monomials left-divisible by $m_1$
and those right-divisible by $m_2$.

\begin{algorithm}
\caption{Homogeneous Non-Commutative Factorization}
\begin{algorithmic}[1]\label{homogeneousAlgorithm}
\REQUIRE A homogeneous polynomial $F$ of degree $n$; \\
the desired degrees $h$ and $k$ for the factorization 
\ENSURE Either a negative answer or the two factors $G$ and $H$ 
of desired degree 
\smallskip
  \STATE ``Select'' any monomial in $F$ and factor it in
          two parts $\hat{G}$ of degree $h$ and $\hat{H}$ of degree $k$.
  \FORALL{monomials $M$ in $F$}
       \IF{$\hat{G}$ left-divides $M$}
         \STATE $H:=H + R$, with $R$ the left-quotient of $M$ by $\hat{G}$
       \ENDIF
       \IF{$\hat{H}$ right-divides $M$}
         \STATE $G:=G + L$, with $L$ the right-quotient of $M$ by $\hat{H}$
       \ENDIF
     \ENDFOR
  \IF{F=G H} 
     \STATE Return $G$ and $H$
  \ELSE
     \STATE Return ``Irreducible''
  \ENDIF
\end{algorithmic}
\end{algorithm}

\begin{remark}
The correctness of Algorithm~\ref{homogeneousAlgorithm} 
implies that for given degrees
there is a unique factorization.
One can prove more:
the factorization is essentially unique, i.e.
if a polynomial $F$ is factored as
$F= G_1 H_1 = G_2 H_2$ with
$\deg G_1 = i < j = \deg G_2$ then
the two factorizations must have a common refinement:
$F = G_1 J H_2$.

This is proved as follows:
we can always assume
\begin{equation}
G_1 = x_1 \cdots x_i + R_1, \; \; \; \; G_2 = x_1 \cdots x_j + R_2.
\end{equation}
\noindent after dividing by a suitable element of the field.

We can now consider the 
polynomial $J_0$ given by the sum of the elements
in 
\begin{equation}
\begin{split}
 & \{ \text{monomial } m \text{ in } G_2 \text{ left-divisible by } x_1 \cdots x_i, \\
       & \text{ i.e. 
of the form } c x_1 \cdots x_i y_1 \cdots y_{j-i} \}.
\end{split}
\end{equation}

\newpage
We have that the set of monomials $m$ in $F$ that
are  left-divisible by $x_1 \cdots x_i$ can be 
obtained as $J_0 H_2$ as well as $x_1 \cdots x_i H_1$.

Therefore we can take the polynomial $J$ given by the sum
of the elements in 
\begin{equation}
\begin{split}
& \{ \text{left-quotient of an element of } J_o \text{ by } x_1 \cdots x_i, \\
& \text{ i.e. a monomial of the form } c y_1 \cdots y_{j-i} \}
\end{split}
\end{equation} 
from which follows
\begin{equation}
x_1 \cdots x_i H_1 = x_1 \cdots x_i J H_2 
\end{equation}
\noindent that implies $H_1 = J H_2$ and $G_2 = G_1 J$.
\end{remark}

\medskip

\section{The General Case}
The general case is more complicated because the 
factorization is not unique anymore even if
we fix the degrees of the factors 
(see Subsection~\ref{example} for a simple example).

\subsection{Exponential Growth of the Number of Factorizations}
Teo Mora (Univ.~of Genoa) has noticed that
for any univariate polynomial $f(t)$ that
is factored 
as $f_1(t) \cdots f_k(t)$ (with all distinct factors),
if we consider $f(XY)$ 
we have the following non-commutative
factorizations for the
polynomial $Y f(XY)$:
\begin{equation}
\begin{split}
Yf(XY) &= Y f_1(XY) \cdots f_k(XY) = \\
f_1(YX) Y \cdots f_k(XY) &= 
f_1(XY) \cdots Y f_k(XY).
\end{split}
\end{equation}
\noindent In such a way Teo Mora proves an exponential lower-bound 
on the number of factorizations with respect to the degree.
However if we homogenize an inhomogeneous
polynomial we are left with just one possible
factorization.
This is explained by the fact that there
are different ways to homogenize.

\begin{example}
Clearly $x^2-1$ is factored as $(x-1)(x+1)$
but it can be homogenized as $x^2-y^2$, which 
is irreducible, or as $x^2-xy+xy-y^2$, which
is factored as $(x-y)(x+y)$, which corresponds
to the factorization $(x-1)(x+1)$.
\end{example}

\begin{remark}\label{zeroDim}
The finiteness of the factorization is still unproved.
A formal proof could be achieved by proving the $0$-dimensionality
of the system produced by the algorithm.
\end{remark}

\newpage
\subsection{The General Algorithm}

Let us consider the problem of factorizing $F$
of degree $n$ as $F=G H$, with $G$ of degree $h$
and $H$ of degree $k$.
The main idea of the algorithm is to use
the relations between the homogeneous parts $F_{n-j}$
(of degree $n-j$) of $F$
and the homogeneous parts $G_{h-j}$ (of degree $h-j$) of $G$,
$H_{k-j}$ (of degree $h-j$) of $H$:
\begin{equation}
\begin{split}
F_n &= G_h H_k \\
F_{n-1} &= G_h H_{k-1} + G_{h-1} H_k \\
F_{n-2} - G_{h-1} H_{k-1} &= G_h H_{k-2} + G_{h-2} H_k\\
F_{n-3} - G_{h-1} H_{k-2} - G_{h-2} H_{k-1} & = G_h H_{k-3} + G_{h-3} H_k \\
& \dots \\
\end{split}
\end{equation}

\noindent It is possible to determine $G_{h-j}$ and $H_{k-j}$ in the right hand side 
by ``inspection'' of the left hand side
(similarly to the homogeneous case, by searching for monomials that have certain ``substrings'').
The main difference from the homogeneous case is that
we must take into account possible cancellations of terms
in the right-hand side, which corresponds
to possible partial overlaps of monomials in $G_h$ and $H_k$.
For each possible cancellation between
$G_h H_{k-j}$ and $G_{h-j} H_k$ we introduce
new ``symbols'', i.e. an extension of our ground field.
The subsequent relations will determine algebraic relations
on the new elements of the field that will produce
a system of polynomial equation, which 
we can solve by a Gr\"obner basis
computation.

\begin{algorithm}
\caption{Non-Commutative Factorization}
\begin{algorithmic}[1]\label{algorithm}
\REQUIRE A polynomial $F$ of degree $n$; \\
the desired degrees $h$ and $k$ for the factorization 
\ENSURE The list of possible factorization of $F$ in $G_h$ and $H_k$
in two parts of degree $h$ and $k$
\smallskip
  \STATE Use Algorithm~\ref{homogeneousAlgorithm} to factorize the homogeneous part of $F$ of highest degree 
in $G_h$ and $H_k$

  \STATE  ``Select'' monomials $\hat{G}= x_1 \dots x_h$ of $G_h$, 
  and $\hat{H}= y_1 \dots y_k$ of $H_k$
  \STATE $\hat{F}_{n-j}:= F_{n-j}$ for all $j=1\dots n$. 
  \FOR{$j$ in $1 \dots n$} 
  \STATE $\hat{F}_{n-j}:= F_{n-j} - \sum_{i=1}^{j-1} G_{h-i} H_{k-j+i}$  
     \IF{the last $j$ variables $x_{h-j+1} \dots x_h$ of $\hat{G}$ 
         are equal to 
         the first $j$ variables $y_1 \dots y_j$ of $\hat{H}$}
     \STATE Consider the coefficient $c$ 
     of $x_1 \dots x_{h-j} y_{j+1} \dots y_k$ in $\hat{F}_{n-j}$
     \STATE $\hat{F}_{n-j}:= 
            \hat{F}_{n-j}- c x_1 \dots x_{h-j} y_{j+1} \dots y_k$
     \STATE $\hat{K}= \hat{K}(\alpha)$ for a new symbol $\alpha$
     \STATE $G_{h-j}:= G_{h-j} + \alpha x_1 \dots x_{h-j} y_{j+1} \dots y_k$
     \STATE $H_{k-j}:= H_{k-j} + 
                       (c-\alpha)x_1 \dots x_{h-j} y_{j+1} \dots y_k$
     \ENDIF
     \FORALL{monomials $M$ in $\hat{F}_{n-j}$}
       \IF{$\hat{G}$ left-divides $M$}
         \STATE $H_{h-j}:=H_{h-j}+ d R$, with $R$ the left-quotient of $M$ by $\hat{G}$
       \ENDIF
       \IF{$\hat{H}$ right-divides $M$}
         \STATE $G_{k-j}:=G_{k-j}+ d L$, with $L$ the right-quotient of $M$ by $\hat{H}$
       \ENDIF
     \ENDFOR
  \ENDFOR
  \STATE Consider $G:= \sum_{i=0}^h G_h$ and $H:= \sum_{i=0}^k H_i$
  \STATE Find the possible values of the new symbols such that $F=G H$
  \STATE Return $G$, $H$ and a description of all possible values of the new symbols
\end{algorithmic}
\end{algorithm}

\begin{remark}
The description of the possible values of the new symbols is in general 
provided by a system of polynomial equations on the new symbols.
Our implementations allows to choose whether
the system should be normalized in the form
of the reduced lexicographic Gr\"obner basis.
\end{remark}

\subsection{One Interesting Simple Example}\label{example}
Let us consider $K=\mathbb{F}_p$, with $p>2$
and $F:= yxyxy-y$.
We can use our algorithm to factorize $F$
in two factors of degree $2$ and $3$.

Our procedure for this example could be summarized as follows
\begin{enumerate}
\item The head $F_n$ of $F$ is just the monomial $yxyxy$
which is factored in $G_2:=yx$ and $H_3:=yxy$. 
\item From
$F_4 = 0 = yx H_2 + G_1 yxy$   
we get $H_2=G_1=0$; 
\item From 
$F_3 - G_1 H_2 = 0 = yx H_1 + G_0 yxy$
we detect a possible cancellation in the right hand side,
which forces us to introduce a new symbol $\alpha$
as possible coefficient, which implies
$H_1:=\alpha y$, $G_0 = -\alpha$; 
\item From
$F_2 - G_1 H_1 - G_0 H_2 = 0 = yx H_0$
we get $H_0=0$; 
\item By using the relation 
$F=yxyxy-y= 
(\sum_{i=0}^2 G_i)(\sum_{j=0}^3 H_j) =
(yx-\alpha)
(yxy+\alpha y)
$
we get $\alpha^2=1$ we gives the two different solutions to
the factorization $(yx \pm 1)(yxy \mp y)$.
\end{enumerate}

\section{Improvements}
The main problem of non-commutative factorization is the
exponential number of cases to be considered.
Our improvements reduce the number of possible
cases to be considered by  
\begin{enumerate}
\item reducing the number of algebraic extensions;
\item reducing the number of possible factorizations
in two factors.
\end{enumerate}

\subsection{Reducing the Extensions}
It is possible to avoid the introduction of 
extensions in the coefficient field by carefully choosing 
$\hat{G}$ and $\hat{H}$ in Algorithm~\ref{algorithm}.
In particular we must choose them in a way
that reduces the number of overlaps between them,
since
for each overlap between the last part of $G$
and the first of $H$ of length $j$ we are forced
to consider a possible cancellation which 
produces a new extension.

\subsection{Commutative Images}
It is possible to immediately detect
some impossible factorization by considering
the commutative version of the given
polynomial and its commutative factors.
If the commutative polynomial
has the same degree as the original one,
then the commutative factors are in general
a refinement of the possible 
non-commutative factorization and therefore
we can greatly limit the number of cases
to be considered. 
A very simplified version of this idea could be
used as follows:
if the commutative version of a given polynomial $F$
can be factored in irreducible factors
of degree $a_1, \dots, a_k$, then 
we need only consider as possible non-commutative
factors those of degrees $b_1$ and $b_2$
where $b_1$ is a solution of
the binary knapsack problem for the $a_i$
and $b_2=\deg F - b_1$ (i.e. $b_1$ is a sum of
some~$a_i$).

This simple approach can be extended
to the other cases expect when
the commutative image is $0$,
by considering the commutative homogeneous
parts of the original polynomial and use them
to reduce the cases.
Another improvement may come
from considering
other quotients of the algebra.

\section{The Axiom Implementation}
We have implemented Algorithm~\ref{algorithm} and its 
improvements in the \texttt{Axiom} computer algebra system.
The choice of the \texttt{Axiom} computer
algebra system is due to the flexibility of this system
and to the fact that \texttt{Axiom} already provides
constructors for non-commutative algebraic structures.
In particular \texttt{MonoidRing} provides
a constructor for polynomials over any monoid
and any coefficient ring, and
\texttt{FreeMonoid} provides one for
free monoids, which is exactly what is needed in our case.
\newpage
\noindent This package is going to be part of standard \texttt{Axiom}
but before this happens you will need to load
the code by typing
\begin{verbatim}
)r daven.input
\end{verbatim}

\noindent The main command is \texttt{NCFactor} which is used as
follows
\begin{verbatim}
NCFactor(y*x*y*x*y-y*x*y);
\end{verbatim}

\noindent which outputs the list of its factorizations
for all possible degrees.

\section{Future Work}
We plan to fully
integrate the non-commutative factorization
in the next version of the \texttt{Axiom} computer algebra system.
There are two open questions we will be working on:
a formal proof (see Remark~\ref{zeroDim})  of the finiteness of factorizations
and the question whether other quotients of the algebra can be used.

\subsection*{Acknowledgments}
We would like to thank James~H. Davenport
for letting us work on his notes.
We would also like to thank
Patrizia Gianni, 
Barry Trager 
for their
help with the \texttt{Axiom} system.

\bibliographystyle{alpha}
\bibliography{non-comm_preprint.bib}

\end{document}